\documentclass[aps,twocolumn,amssymb,amsfonts,floatfix,superscriptaddress,prl,longbibliography]{revtex4-1}
\pdfoutput=1
\usepackage[applemac]{inputenc}
\usepackage[english]{babel}
\usepackage[pdftex]{graphicx}
\usepackage{color}
\usepackage{ulem}
\usepackage{multirow}
\usepackage{physics}

\usepackage{dcolumn}
\usepackage{bm}
\usepackage{mathbbol}

\usepackage{amsmath,amsthm}
\usepackage{times}
\usepackage{hyperref}

\DeclareMathAlphabet\bmcal{OMS}{cmsy}{b}{n}

\begin{document}
\title{Positive quantum Lyapunov exponents in experimental systems with a regular classical limit}
\author{Sa\'ul Pilatowsky-Cameo}
\affiliation{Instituto de Ciencias Nucleares, Universidad Nacional Aut\'onoma de M\'exico, Apdo. Postal 70-543, C.P. 04510  CDMX, Mexico}
\author{Jorge Ch\'avez-Carlos}
\affiliation{Instituto de Ciencias Nucleares, Universidad Nacional Aut\'onoma de M\'exico, Apdo. Postal 70-543, C.P. 04510  CDMX, Mexico}
\author{Miguel A. Bastarrachea-Magnani}
\affiliation{Department of Physics and Astronomy, Aarhus University, Ny Munkegade, DK-8000 Aarhus C, Denmark}
\author{Pavel Str\'ansk\'y}
\affiliation{Faculty of Mathematics and Physics, Charles University, V Hole\v{s}ovi\v{c}k\'ach 2, Prague 180 00, Czech Republic}
\author{Sergio Lerma-Hern\'andez}
\affiliation{Facultad  de F\'\i sica, Universidad Veracruzana, Circuito Aguirre Beltr\'an s/n, C.P. 91000  Xalapa, Mexico}
\author{Lea F. Santos}
\affiliation{Department of Physics, Yeshiva University, New York, New York 10016, USA} 
\author{Jorge G. Hirsch} 
\affiliation{Instituto de Ciencias Nucleares, Universidad Nacional Aut\'onoma de M\'exico, Apdo. Postal 70-543, C.P. 04510  CDMX, Mexico}

%%%%%%%%%%%%%%%%%%%%%%%%%%%%%%%%%%%%%%%%

\begin{abstract}
Quantum chaos refers to signatures of classical chaos found in the quantum domain. Recently, it has become common to equate the exponential behavior of out-of-time order correlators (OTOCs) with quantum chaos. The quantum-classical correspondence between the OTOC exponential growth and chaos in the classical limit has indeed been corroborated theoretically for some systems and there are several projects to do the same experimentally. The Dicke model, in particular, which has a regular and a chaotic regime, is currently under intense investigation by experiments with trapped ions. We show, however, that for experimentally accessible parameters, OTOCs can grow exponentially also when the Dicke model is in the regular regime. The same holds for the Lipkin-Meshkov-Glick model, which is integrable and also experimentally realizable. The exponential behavior in these cases are due to unstable stationary points, not to chaos.  
\end{abstract} 

\maketitle

%%%%%%%%%%%%%%%% INTRODUCTION %%%%%%%%%%%%%%%%%

Classical chaos in Hamiltonian systems is typically defined by means of the sensitive dependence on initial conditions, which leads to positive Lyapunov exponents (LEs) \cite{OttBook}. But this alone is not a complete definition of chaos. Consider, for example, the simple pendulum. Its upright position corresponds to a stationary point that is unstable. It has a positive LE, as any genuine chaotic system, although it is completely integrable. The pendulum does not exhibit chaotic behaviors, such as non-periodicity and mixing~\cite{Gaspard1998}. Its unstable point and the phase-space orbits emanating from it have measure zero with respect to the rest of the phase space. In this work, we investigate what happens to such unstable points in the quantum domain. 

It was argued in~\cite{RozenbaumARXIV2019} that quantum mechanics can bring chaos to classical systems that are non-chaotic. This idea was inspired by Ref.~\cite{Bunimovich2019}, where a standard non-chaotic classical billiard became chaotic when the point particle was substituted by a finite-size hard sphere. By making a parallel between the semiclassical dynamics of a quantum wave packet and the motion of a finite-size classical particle, it was shown in~\cite{RozenbaumARXIV2019} that quantum chaos can emerge in regular classical billiards. Quantum chaos in this case refers to the exponentially fast growth of the out-of-time ordered correlator (OTOC) at short times.

The OTOC quantifies the degree of noncommutativity in time between two  operators. It was introduced in the context of superconductivity~\cite{Larkin1969} to measure the instability of the trajectories of electrons scattered by impurities. Recently, the OTOC became a key quantity in definitions of many-body quantum chaos~\cite{KitaevTalk,Maldacena2016PRD,Maldacena2016JHEP,Roberts2015PRL,Fan2017,Luitz2017b,Torres2018,Borgonovi2019,Yan2019}, analysis of the quantum-classical correspondence of chaotic systems~\cite{Rozenbaum2017,Rozenbaum2019,Hashimoto2017,Chavez2019,Garcia2018,Jalabert2018,Fortes2019,Rammensee2018,Lakshminarayan2019}, and studies of the scrambling of quantum information~\cite{Garttner2018,Lewis-Swan2019} and quantum phase transitions~\cite{Shen2017,Wang2019B}. The OTOC has been measured experimentally with ion traps~\cite{Garttner2017} and nuclear magnetic resonance platforms~\cite{Li2017,Wei2018,NiknamARXIV}.

Depending on how the OTOC is computed, it may be called microcanonical OTOC (MOTOC)~\cite{Hashimoto2017}, fidelity OTOC (FOTOC)~\cite{Lewis-Swan2019}, thermal OTOC~\cite{Maldacena2016JHEP}, and OTOC for specific initial states~\cite{Rozenbaum2017,RozenbaumARXIV2019}. The exponential growth rate of the latter, of the MOTOC~\cite{Chavez2019}, and of the FOTOC~\cite{Lewis-Swan2019} was shown to be related with the classical LE of chaotic systems. This has justified referring to the OTOCs exponential growth rates as quantum LEs and associating their exponential behavior with the notion of quantum chaos. 

However, based on a semiclassical quantization approach, it was recently shown that, in general, the OTOC can grow exponentially fast also in one-degree-of-freedom quantum systems that are not globally chaotic, but are critical~\cite{Hummel2019}. Here, we show that this happens also for the Dicke model, which has two degrees of freedom and is used to describe strongly interacting light-matter systems~\cite{Dicke1954,Garraway2011,Kirton2019}.  The Dicke model presents chaotic and regular regimes and is of great experimental relevance. It has been realized experimentally with cold atoms~\cite{Baumann2010,Baumann2011,Ritsch2013,Klinder2015}, by means of cavity Raman transitions~\cite{Baden2014,Zhang2018}, and with ion traps~\cite{Safavi2018}. We study the FOTOC, because this quantity is directly measured by trapped ion experiments, and consider parameters and initial states used in these experiments.

The Dicke model has unstable points that give rise to positive LEs in the regular regime. These points and the orbits emanating from them have measure zero~\cite{footChaos}. In the quantum domain, on the other hand, we find that the FOTOC grows exponentially not only for initial states centered at the classically unstable point, but also for generic states centered at the surrounding points with zero classical LEs. Quantum mechanics therefore generates instability in a region where the classical dynamics is stable. Following the current terminology, we then refer to these regions as ``quantum chaotic'', although one may ponder whether, similarly to the above discussion about classical chaos, additional conditions, on top of the exponential growth of the OTOCs, are needed for defining quantum chaos.

The OTOC grows exponentially also at the critical point of the Lipkin-Meshkov-Glick (LMG) model~\cite{Pappalardi2018}. This is a one-degree of freedom classically integrable system introduced in nuclear physics~\cite{Lipkin1965a,*Lipkin1965b,*Lipkin1965c} and realized experimentally with cold atoms~\cite{Gross2010,Zibold2010}  and nuclear magnetic resonance platforms~\cite{Ferreira2013}.  By studying the FOTOC, we show that the exponential behavior persists in the vicinity of this critical point as well.

%%%%%%%%%%%%%%%%%%%%%%%%%%%%%%%%%%%%%%%%%%%%%%%%
{\it The unstable points of the LMG and Dicke models.} 
In a classical Hamiltonian system with real first-order differential equations $d\bm{x}/dt=\bmcal{F}(\bm{x})$, where $\bm{x}=(\bm{q},\bm{p})$ are the generalized coordinates and momenta,  a point $\bm{x} = \bm{x}_0$ is stationary when $\bmcal{F}(\bm{x}_0)=0$. This point is unstable when  at least one of the positive-negative pairs of eigenvalues of the Jacobian matrix of  $\bmcal{F}$ evaluated at $\bm{x}_0$ has a nonzero real part. The LE of this point equals the maximum of these real part values [see the Supplemental Material (SM) in~\cite{SM} for more details]. Both the LMG and the Dicke model in the classical limit present stationary  points with positive LEs.
 
The LMG model~\cite{Lipkin1965a,*Lipkin1965b,*Lipkin1965c} describes the collective motion of a set of $N$ two-level systems mutually interacting. Its quantum Hamiltonian is given by 
\begin{equation}
	\hat{H}_{\text{LMG}}= \Omega \hat{J}_z + \frac{2\xi}{N} \hat{J}_x^2,
\end{equation}
where $\hbar=1$,  $\Omega$ is the energy difference of the two-level systems, $\xi$ is the coupling strength, $\hat{J}_{x,y,z}=(1/2)\sum_{n=1}^N \sigma_{x,y,z}^{(n)} $ are the collective pseudospin operators given by the sum of Pauli matrices  $\sigma_{x,y,z}^{(n)} $  for each two-level system $n$, and $j=N/2$ gives the size of the system, with $j(j+1)$ being the eigenvalue of the total spin operator $\bm{\hat{J}}^2 =  \hat{J}_x^2 + \hat{J}_y^2 + \hat{J}_z^2$. This model has been employed, for example, in studies of ground state quantum phase transitions (QPTs) and excited state quantum phase transitions (ESQPTs)~\cite{Caprio2008,Cejnar2010,SantosBernal2015,Santos2016},  entanglement~\cite{Vidal2004,Dusuel2004}, and quantum speed limit~\cite{Wang2019}. 

The classical LMG Hamiltonian is obtained by taking the expectation value of $\hat{H}_{\text{LMG}} /j $ on Bloch coherent states $ |z\rangle=\left(1+\left|z\right|^{2}\right)^{-j} e^{z \hat{J}_+}|j, -j\rangle$, where $|j, -j\rangle$ is the state with the lowest pseudospin projection and $\hat{J}_+$ is the raising operator. Defining $z$ in terms of the canonical variables $(Q,P)$ as $z=(Q-iP)/\sqrt{4 -(Q^2+P^2)}$  and neglecting $O(1/j)$ terms, the classical LMG Hamiltonian reads~
\begin{equation}
 H_{\text{LMG}}(Q,P) \!=\! \frac{\Omega}{2} \left(Q^2 \!+\! P^2\right)-\Omega+
   \xi \left(Q^2 \!-\! \frac{Q^2 P^2}{4} \!-\! \frac{Q^4}{4}\right). 
   \label{Eq:LMGclassical}
\end{equation}

\begin{figure}[ht!]
\centering
\begin{tabular}{cc}
\includegraphics[width= 0.4 \columnwidth]{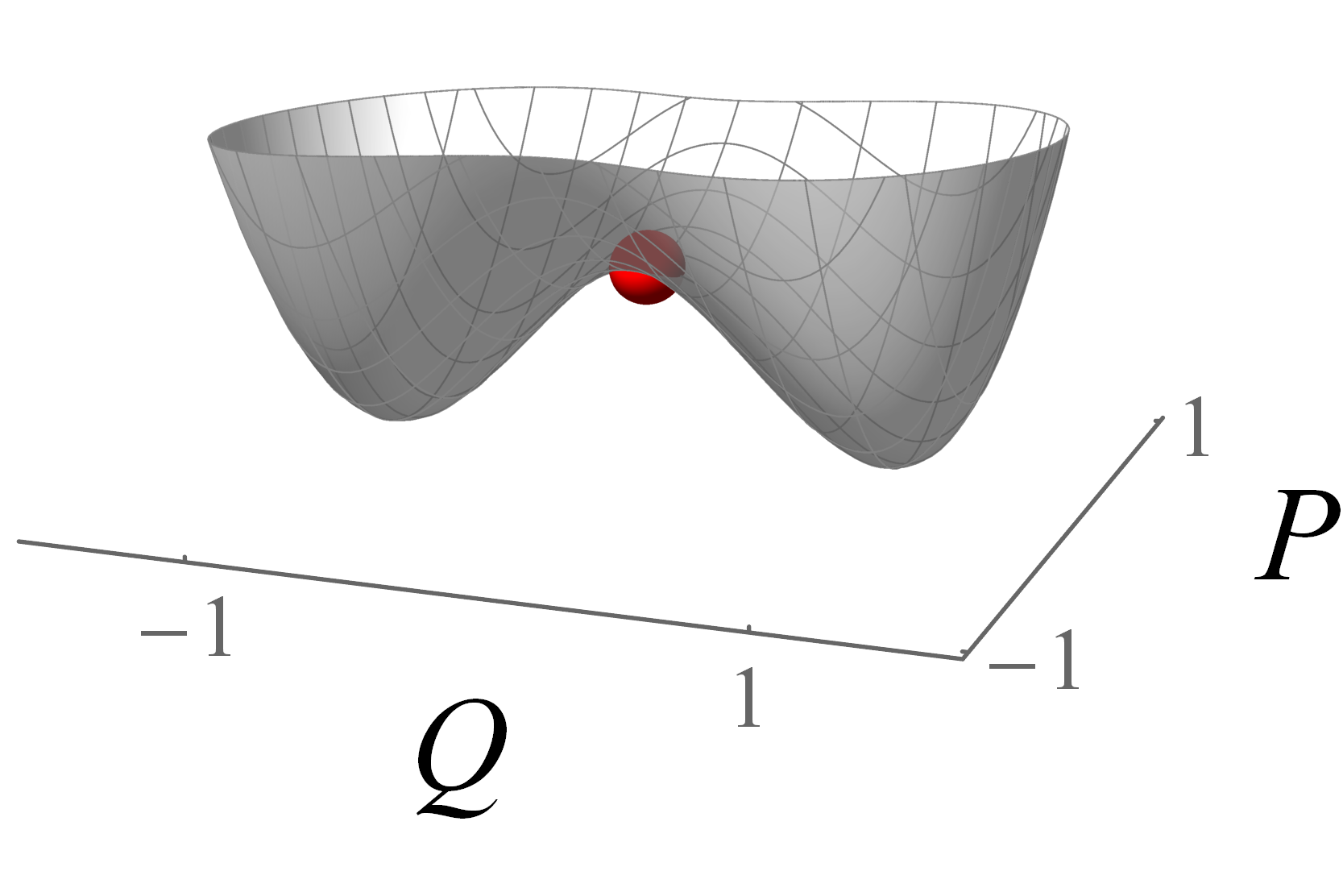}  
& \includegraphics[width= 0.4 \columnwidth]{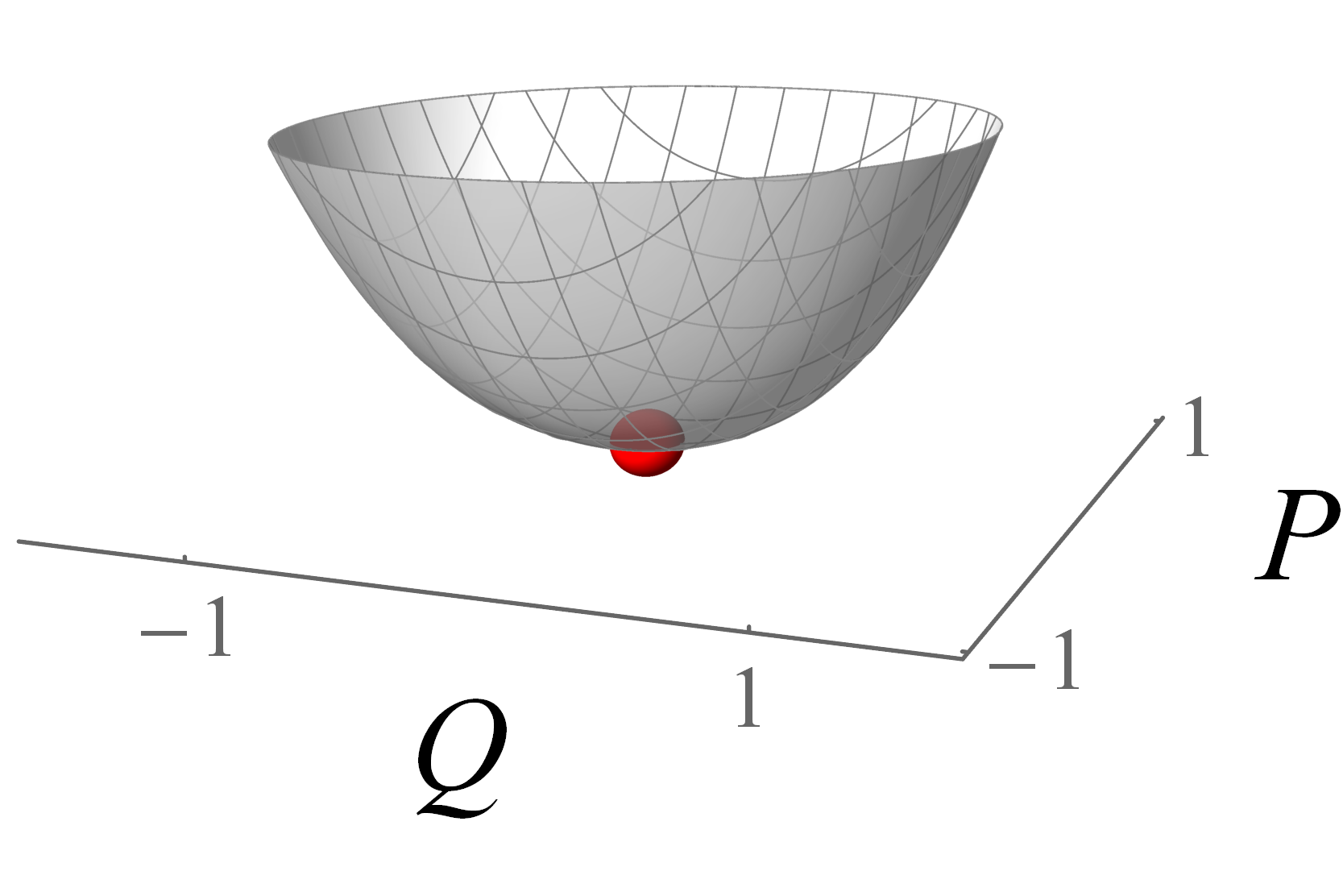} \\
(a) $\Omega = 1$  & (b) $\Omega = 3$ 
\\
\\
\multicolumn{2}{c}{\includegraphics[width= 0.96 \columnwidth]{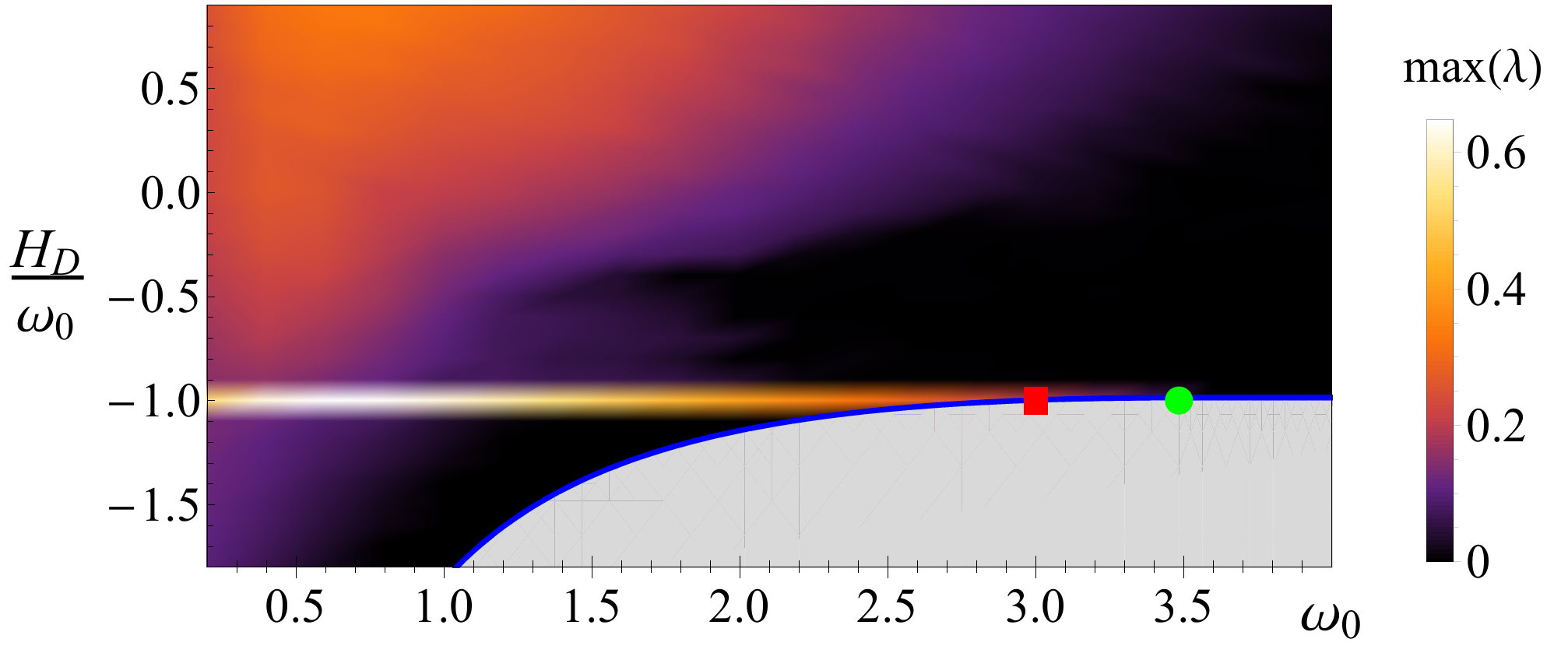} }\\
\multicolumn{2}{c}{(c)}
\end{tabular} 
\caption{Top: Energy surface for the classical LMG model for two values of the parameter $\Omega$  fixing $\xi=-1$. The stationary point ${\bm{x}_0 = (Q=0,P=0)}$ is marked with a red sphere. It is a saddle point for  $\Omega=1$ (a) and a minimum for ${\Omega=3}$ (b). Panel (c): Each colored point corresponds to the maximal classical Lyapunov exponent for the Dicke model in a plane resulting from the intersection of an energy shell (with energy indicated by the vertical axis) and the hyperplane ${p=0}$. This is done for different values of $\omega_0$ as indicated by the horizontal axis. We fix $\gamma=0.66$ and $\omega=0.5$. The red square at $\omega_0=3$ is the unstable point studied in Fig.~\ref{Fig:FotocExploration}. The green circle at $\omega_{0c}=3.48$ is the critical point that marks the ground state quantum phase transition.
}
 \label{Fig01}
 \end{figure}

Hamiltonian (\ref{Eq:LMGclassical}) is regular, but its stationary point ${\bm{x}_0 = (Q=0,P=0)}$ is unstable and presents a positive LE given by 
\begin{equation}
\lambda = \sqrt{-(\Omega^2+2\Omega \xi)}
\label{Eq:Lambda_LMG}
\end{equation} 
when $\Omega<-2\xi$ (see \cite{Pappalardi2018} and SM~\cite{SM}). Figures~\ref{Fig01}(a) and \ref{Fig01}(b) show the energy surface of the classical LMG model for $\xi=-1$ and two values of $\Omega$. When ${ \Omega \geq -2\xi}$, $\bm{x}_0$ is a minimum, while for ${\Omega < -2\xi}$, $\bm{x}_0$ becomes a saddle point and is therefore unstable.

In the quantum domain, this saddle point is associated with an ESQPT. A main signature of ESQPTs is the divergence of the density of states at an energy denoted by $E_{\text{ESQPT}}$. In the mean-field approximation, it has been shown that this energy coincides with the energy of the classical system at the saddle point~\cite{Cejnar2006,Caprio2008}, that is, for the LMG model, $E_{\text{ESQPT}}^{\text{LMG}}/ j = H_{\text{LMG}}(\bm{x}_0)=-\Omega$. 

%-------------------------------------------------------------------
 %-------------------------------------------------------------------

The Dicke model is a collection of $N$  two-level atoms of level spacing $\omega_0$ coupled to a quantized radiation field of frequency $\omega$. The Hamiltonian is given by
\begin{equation}
\hat{H}_{D}=\dfrac{\omega}{2}(\hat{q}^2+\hat{p}^2)+\omega_{0}\hat{J}_{z}+2 \sqrt{2} \frac{\gamma}{\sqrt{N}} \hat{J}_{x} \, \hat{q}- \frac{\omega}{2},
\label{eq:HD}
\end{equation}
where $\hat{q}=(\hat{a}^{\dagger} + \hat{a})/\sqrt{2}$ and $\hat{p}=i(\hat{a}^{\dagger} - \hat{a})/\sqrt{2}$, with $\hat{a} (\hat{a}^{\dagger})$ being the annihilation (creation) operator,  and $\gamma$ is the atom-field interaction strength. As in the LMG model, in the symmetric atomic subspace, $j=N/2$.

The Dicke model was first used to explain the collective phenomenon of superradiance~\cite{Dicke1954,Hepp1973,*Wang1973,*Carmichael1973}. It is now used  in studies of QPTs and ESQPTs~\cite{Hepp1973,*Wang1973,*Carmichael1973,Castanos2005,Fernandez2011b,Brandes2013,Bastarrachea2014a,Larson2017},  quantum chaos~\cite{Lewenkopf1991,Emary2003PRL,*Emary2003,Bastarrachea2014b,*Bastarrachea2015,*Bastarrachea2016PRE,Chavez2016}, monodromy~\cite{Babelon2009,Kloc2017JPA}, entanglement creation~\cite{Schneider2002,*Lambert2004,*Kloc2017}, nonequilibrium dynamics~\cite{Fernandez2011,Altland2012PRL,Lerma2018,Lerma2019,Kloc2018}, 
OTOC behavior~\cite{Alavirad2019,Lewis-Swan2019}, and quantum batteries~\cite{Andolina2019}.

The classical Dicke Hamiltonian~\cite{Ribeiro2006,Bakemeier2013,Chavez2016} is obtained by taking the expectation value of $\hat{H}_{D}/j$ between the product of Bloch coherent states and Glauber coherent states $|\alpha\rangle=e^{-|\alpha|^2/2}e^{\alpha \hat{a}^\dagger}|0\rangle ,$ where  $\alpha=\sqrt{j/2}(q+ip)\in \mathbb{C}$, and $|0\rangle$ is the photon vacuum. In terms of the canonical variables $(Q,P)$ for the pseudospin and $(q,p)$ for the field~\cite{SM}, it reads 
\begin{equation}
\begin{split}
H_D= \frac{\omega}{2}\left(q^{2}+p^{2}\right) -  \omega_0 +  \frac{\omega_0}{2}\left(Q^{2}+P^{2}\right) \\
+ 2  \gamma\sqrt{1- \frac{1}{4}\left(Q^{2}+P^{2}\right)} \,q \,Q .
\end{split}
\label{Eq:haclPQ}
\end{equation}

The stationary point of the Dicke model is $\bm{x}_0 = (q=0,p=0,Q=0,P=0)$. The LE associated with it can be calculated in terms of $\omega$, $\omega_0$, and $\gamma$, as (see SM~\cite{SM})
\begin{equation}
\lambda=\frac{1}{\sqrt{2}} \sqrt{-\left(\omega^2 + \omega_0^2 \right) + \sqrt{\left(\omega^2 - \omega_0^2 \right)^2+
  16 \gamma^2 \omega \omega_0}}.
\label{Eq:Lambda_D}
\end{equation}
When $\omega_0 < \omega_{0c}= 4 \gamma^2/\omega$, this equation gives a positive value for the LE and the stationary point is unstable. When $\omega_0 > \omega_{0c}$, Eq.~(\ref{Eq:Lambda_D}) has pure imaginary values and the LE is zero.  The critical point $\omega_{0c}$  marks the ground state QPT of the Dicke model. For $\omega_0 < \omega_{0c}$, the system is in the superradiant phase, and for $\omega_0 > \omega_{0c}$, it is in the normal phase. The unstable point is therefore in the superradiant phase. 

Energy surfaces similar to those in Figs.~\ref{Fig01}(a) and \ref{Fig01}(b) can also be drawn for the Dicke model, but in higher dimension. The saddle point of this model is also associated with an ESQPT~\cite{Bastarrachea2014a}, which happens at $E_{\text{ESQPT}}^{\text{D}}/ j= H_D(\bm{x}_0) =- \omega_0$. We stress that, contrary to common belief, the ESQPT in the Dicke model is not directly related with the transition to classical chaos \cite{Chavez2016,Relano2016EPL}.
  
In Fig.~\ref{Fig01}~(c), we show the largest LEs of the Dicke model as functions of the classical excitation energy $H_D/\omega_0$ and of the atomic frequency $\omega_0$, for $\gamma=0.66$ and $\omega=0.5$. Employing frequency units of kHz/$2\pi$, these values coincide with those used in the experiment with ion traps~\cite{Lewis-Swan2019,Safavi2018}. The blue line in the figure depicts the ground state energy and the gray area under it is forbidden. The color gradient indicates the presence or absence of chaos: black represents regular regions and light areas have large LEs. The bright horizontal line at the ESQPT, $H_D/\omega_0=-1$, indicates very large LEs and reflects the instability. 

According to Eq.~(\ref{Eq:Lambda_D}), the maximum LE is obtained for $\omega_0=0.649$, which is approximately the value used in~\cite{Lewis-Swan2019}. As one sees in Fig.~\ref{Fig01}~(c), this classical instability is immersed in a chaotic region of the phase space with positive LEs, so we show some results for it only in the SM~\cite{SM}. Here, our main focus is on the unstable point at $\omega_0=3$, which is marked in the figure with a red square. The phase-space region surrounding this unstable point is regular, with zero LEs everywhere, except for the phase-space orbits emanating from it~\cite{Chavez2016,footRegular}. This is the unstable point that we use in our studies in Fig.~\ref{Fig:FotocExploration}. But before showing those results, let us describe how the quantum and classical evolutions are carried out and compared.

%%%%%%%%%%%%%%%%%%%%%%%%%%%%%%%%%%%%%%%%%%%%%%%%
%%%%%%%%%%%%%%%%%%%%%%%%%%%%%%%%%%%%%%%%%%%%%%%%
{\it Quantum-classical correspondence.--} The OTOC measures the degree of non-commutativity in time between operators $\hat{W}$ and $\hat{V}$,
$O_{toc} (t)= -\langle  \left[\hat{W}(t),\hat{V}(0) \right]^2   \rangle $. It is known as FOTOC when
$\hat{W}=e^{i\delta \phi \hat{G}}$, where $\hat{G}$ is a Hermitian operator and $\delta \phi$ is a small perturbation, and $\hat{V}=| \Psi_0\rangle \langle  \Psi_0|$ is the projection operator onto the initial state. In the perturbative limit, $\delta \phi \ll 1$, the dynamics of the FOTOC agrees with that of the variance of $\hat{G}$ (see~\cite{Lewis-Swan2019} and SM~\cite{SM}),
\begin{equation}
\sigma^2_{G}(t)=\langle \hat{G}^2(t) \rangle- \langle \hat{G}(t)\rangle^2,
\end{equation}
so we refer to this variance as FOTOC and denote its exponential growth rate by $2\Lambda$. In what follows, we refer to $\Lambda$ as the quantum LE. 

The FOTOC enables a direct visualization of the quantum evolution in terms of the dynamics in phase space. It measures the spread of the size of the wave packet and can thus be compared with the variance of the canonical variables in phase space. 

\begin{figure}[h!]
\centering
\begin{tabular}{cc}
\includegraphics[width= 0.48 \columnwidth]{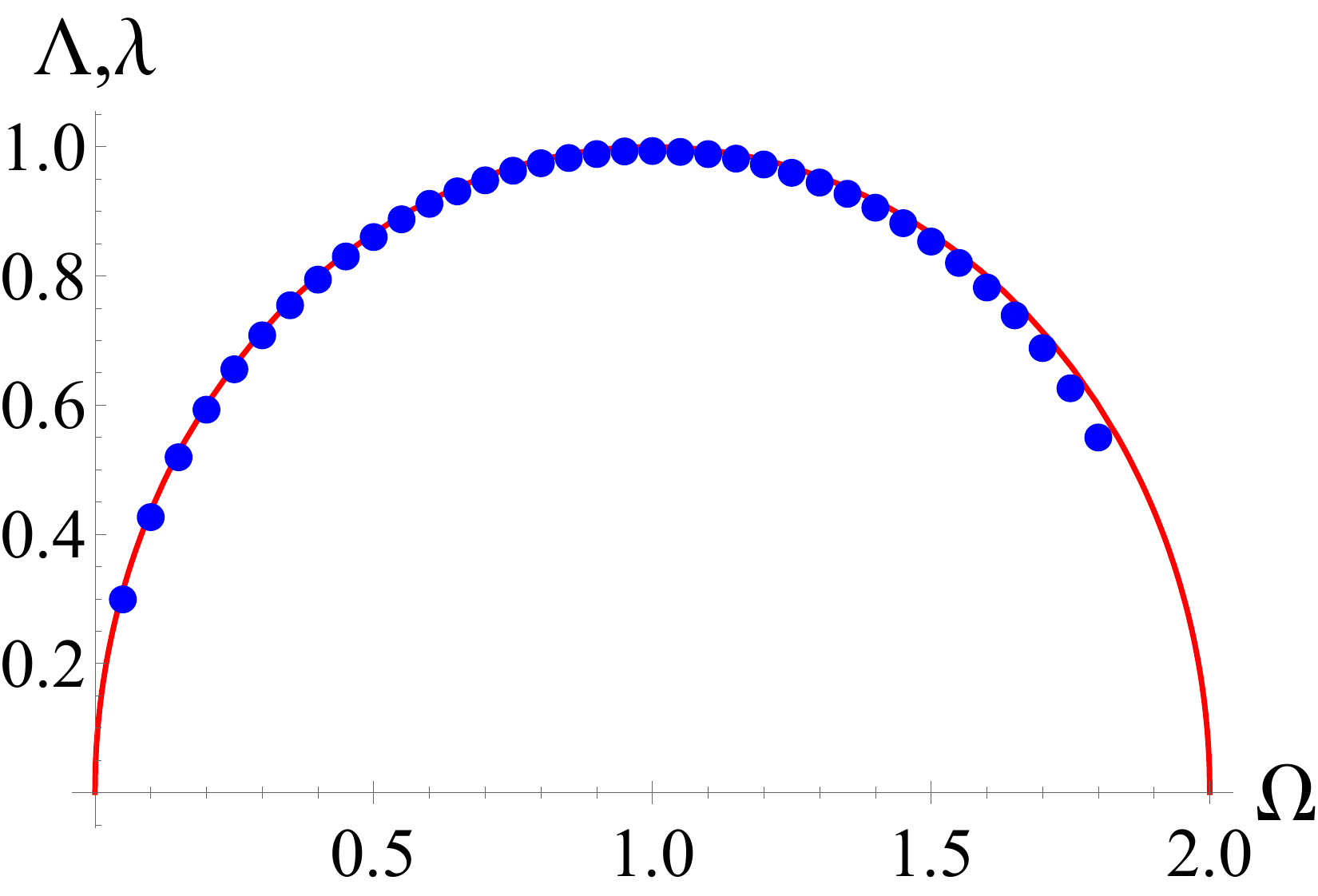}   &
\includegraphics[width= 0.48 \columnwidth]{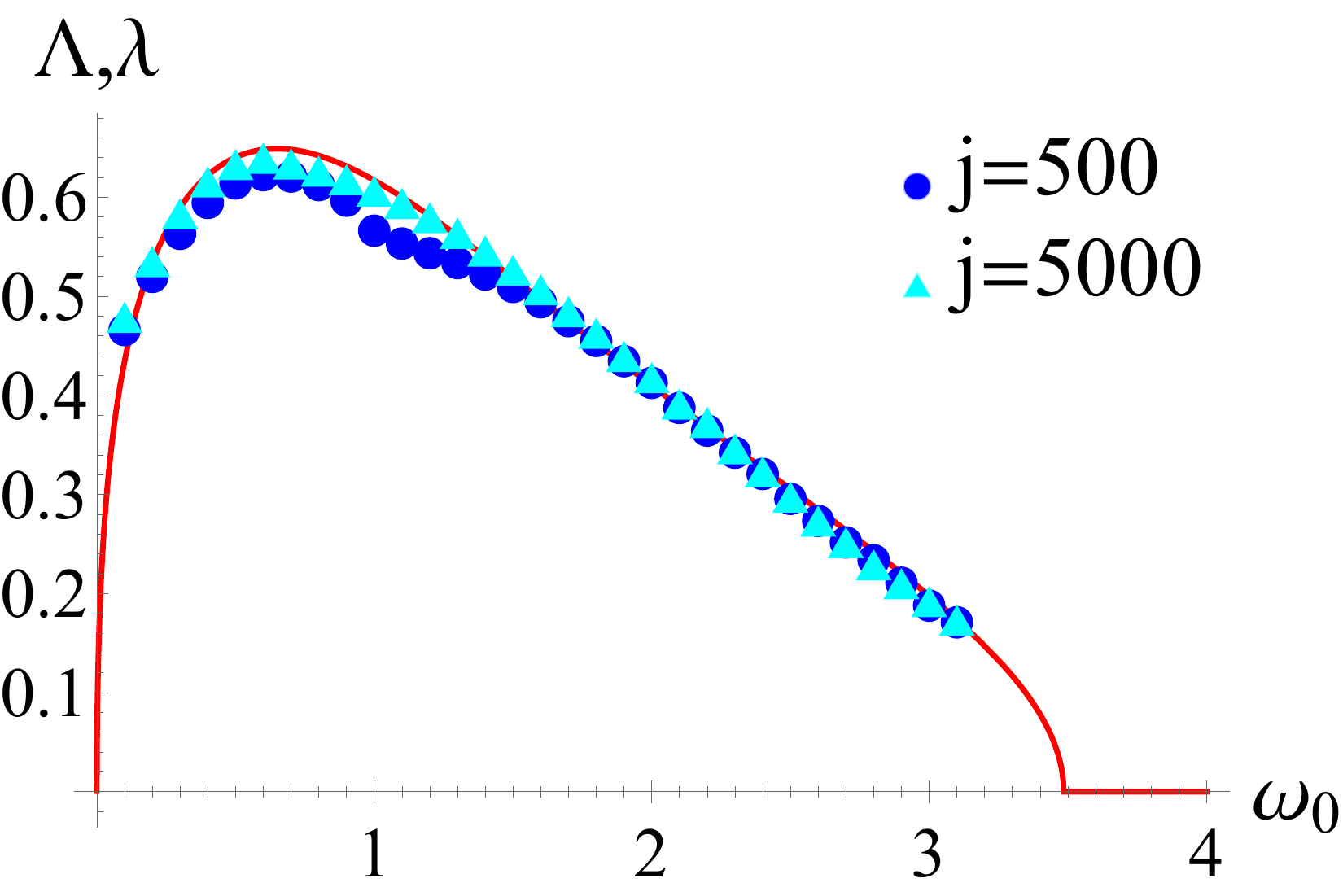} \\
(a) LMG model & (b) Dicke model
\end{tabular} 
\caption{The classical LE $\lambda$ (solid line) and the quantum LE $\Lambda$ (symbols)  for the LMG (a) and the Dicke  (b) model at the unstable point. The results for $\Lambda$ for the LMG model are obtained with the exact quantum evolution and for the Dicke model, the TWA is used. For the LMG model, the FOTOC corresponds to $\sigma_{Q}^2 (t) + \sigma_{P}^2 (t)$, $\xi=-1$, and $j=500$. For the Dicke model, the FOTOC is $\sigma_{Q}^2 (t)+ \sigma_{P}^2 (t)+ \sigma_{q}^2 (t)+ \sigma_{p}^2 (t)$, $\omega=0.5$, $\gamma=0.66$, and the $j$'s are indicated.
}
 \label{Fig:Lambda}
\end{figure}

\begin{figure}[ht]
\centering
\begin{tabular}{cc}
\includegraphics[width= 0.43 \columnwidth]{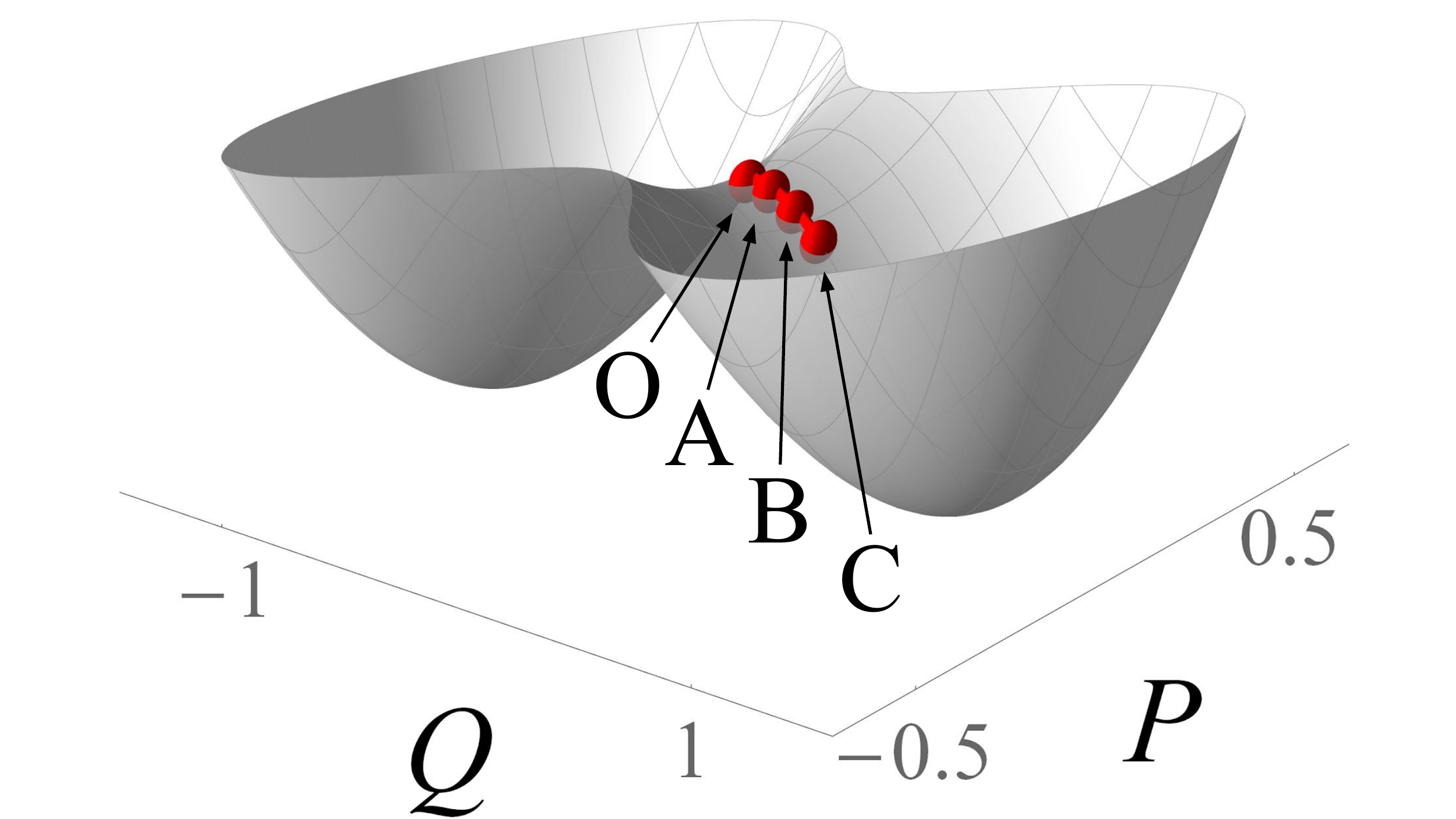} &
\includegraphics[width= 0.53 \columnwidth]{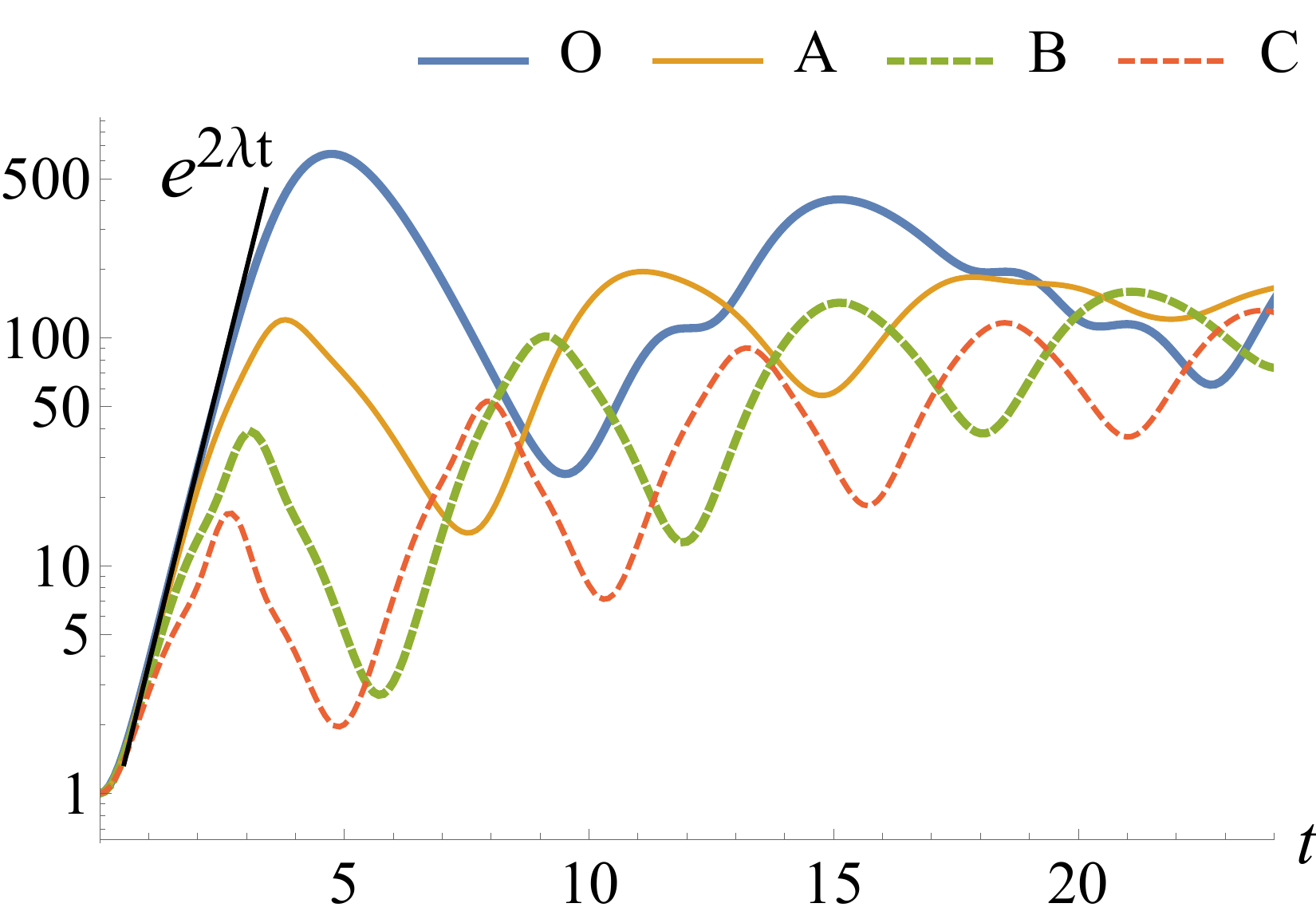} \\
(a) & (b)
\\
\includegraphics[width= 0.43 \columnwidth]{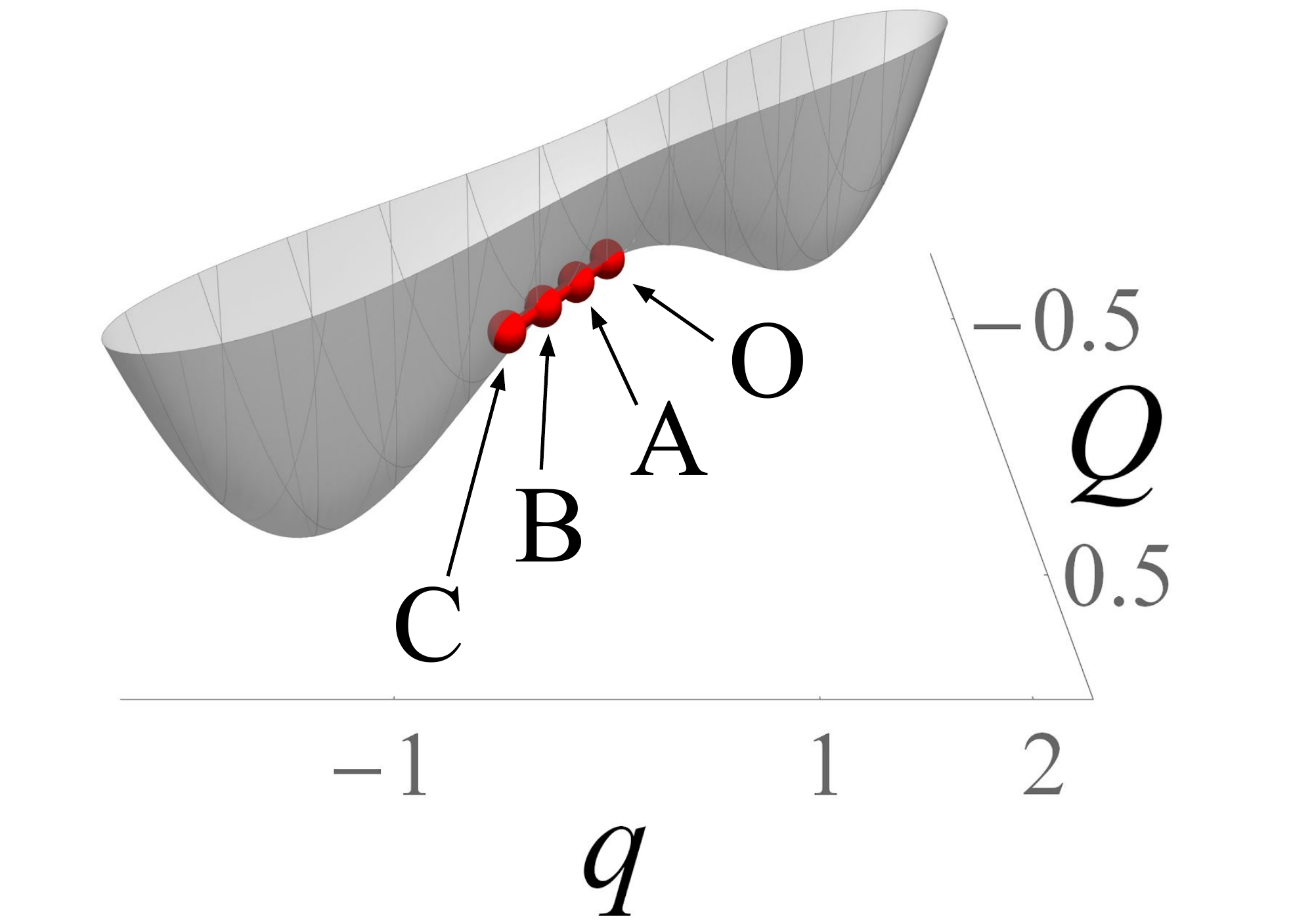} &
\includegraphics[width= 0.53 \columnwidth]{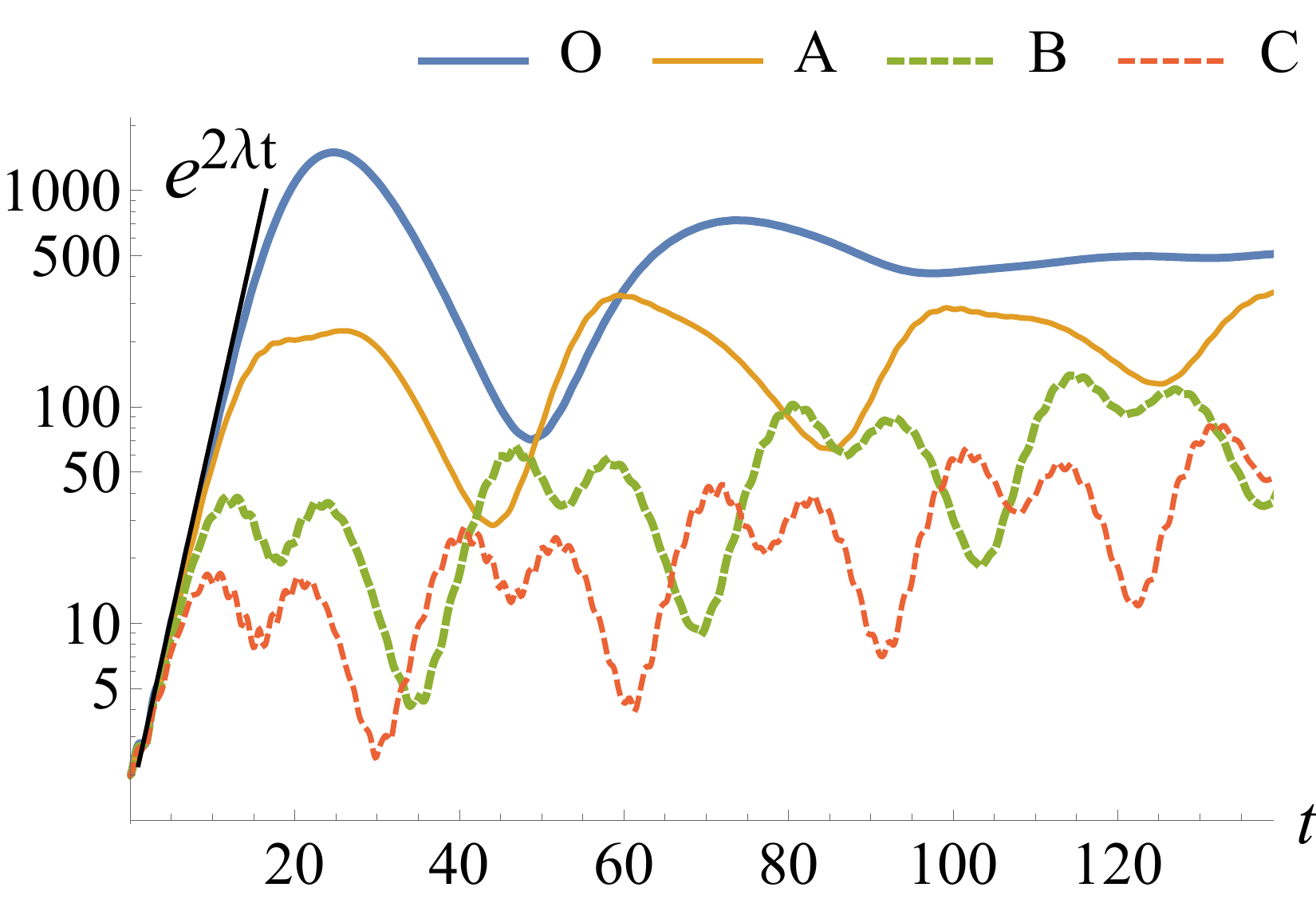} \\
(c) & (d)
\end{tabular} 
\caption{Energy surface of the LMG model (a) and of the Dicke model with $P=p=0$ (c). FOTOC $\sigma_{Q}^2 (t)+ \sigma_{P}^2 (t)$ for the LMG model (b) and FOTOC $\sigma_{Q}^2 (t)+ \sigma_{P}^2 (t) +\sigma_{q}^2 (t) + \sigma_{p}^2 (t)$ for the Dicke model (d). The FOTOC is computed for coherent states centered at the unstable point O and around it, at points A, B, and C. The (black) straight line in (b) and (d) corresponds to the exponential curve with rate given by twice the classical LE. The initial growth rate of the FOTOC for all points and for both models is $2 \Lambda \approx 2 \lambda$. For the LMG model:  $\xi=-1$, $\Omega=1$, $j=500$. The points A, B, and C have constant $P=0$ and $Q=0.1$,$0.2$, and $0.3$, respectively. For the Dicke model: $\omega=0.5$, $\gamma=0.66$, $\omega_0=3$, and $j=500$. The points A, B, and C have $P=p=0$, $Q=0.1, 0.2$, and $0.3$, respectively, and $q$ is chosen so that $H_D=-\omega_0$ for all four points. 
}
\label{Fig:FotocExploration}
\end{figure}

To compute the FOTOC, we consider  initial Bloch coherent states for the LMG model, and  initial products of Bloch and Glauber coherent states for the Dicke model. 
In Fig.~\ref{Fig:Lambda}, we compare the quantum LE obtained for the FOTOC with the classical LE for the LMG (a) and the Dicke (b) model at an unstable point. For the LMG model, the quantum evolution is done exactly. Since the wave packet spreads in both directions in phase space, we analyze the growth of $\sigma^2_{Q}(t) + \sigma^2_{P}(t)$. The agreement between $\lambda$ from Eq.~(\ref{Eq:Lambda_LMG}) and $\Lambda$ is perfect.

%-------------------------------------------------------------------
%-------------------------------------------------------------------
%  

A great advantage of the FOTOC is that it can be computed with semiclassical phase-space methods, such as the truncated Wigner approximation (TWA) \cite{Steel1998,Steel1998,Polkovnikov2010,Schachenmayer2015,Schmitt2019}, 
which makes accessible system sizes that are not achievable with exact diagonalization. This is particularly useful for the Dicke model, which is nonintegrable and where the number of bosons in the field is not limited.

The basic idea of the TWA~\cite{Polkovnikov2010} is to compute the dynamics using the classical equations of motion, but averaging the observable over  a large sample of initial conditions and replacing the classical probability distribution with the Wigner function~\cite{Wigner1932} and the classical observable with the Weyl symbol of the corresponding quantum operator~\cite{Weyl1927}. The random sampling reproduces the quantum fluctuations of a quantum initial state. 

The FOTOC that we study for the Dicke model is $\sigma^2_{Q}(t) + \sigma^2_{P}(t) + \sigma^2_{q}(t) + \sigma^2_{p}(t)$. Employing an efficient basis for the convergence of the eigenstates \cite{Bastarrachea2014PSa}, we evaluate the exact quantum evolution for $j=100$, where the  truncated Hilbert space has 24\,453 converged eigenstates. We verify that for this size, which is already large for exact diagonalization, the exact quantum evolution and the evolution done with the TWA  agree extremely well from $t=0$ up to  times beyond the exponential growth of the FOTOC (see SM~\cite{SM}). This assures us that we can use the TWA to calculate $\Lambda$ for larger $j$'s. For coherent states, the initial Wigner functions are positive and approximately given by normal distributions. Our sampling is done by means of a Monte Carlo method~\cite{Schachenmayer2015} over $\sim 10^4$ random points (see details in SM~\cite{SM}).  As one increases $j$ from $500$ to $5000$, the agreement between $\lambda$ from Eq.~(\ref{Eq:Lambda_D}) and the quantum LE improves, as seen in Fig.~\ref{Fig:Lambda}~(b). 

For our set of parameters, the Dicke model can be separated in fast and slow modes at the ESQPT energy~\cite{Magnani2017}. For the slow mode, an ESQPT as in an effective one-degree-of-freedom Hamiltonian emerges. This confirms the conjecture in~\cite{Hummel2019} that their results might apply also to models with more than one degree of freedom.

%%%%%%%%%%%%%%%%%%%%%%%%%%%%%%%%%%%%%%%%%%%%%%%%
%%%%%%%%%%%%%%%%%%%%%%%%%%%%%%%%%%%%%%%%%%%%%%%%
{\it Quantum activation of the instability.} The results above make evident that, despite the regularity of the systems, both classical and quantum LEs coincide and are positive at the unstable points. We now investigate what happens at the vicinity of the unstable point of the LMG model with $\Omega=1$ and of the Dicke model with $\omega_0=3$. Classically, the LEs in these surrounding regions, in orbits not asymptotically going to or coming from the unstable point, are zero. To analyze what happens in the quantum domain, we study the behavior of the FOTOC as one moves away from the unstable point. 

%-------------------------------------------
%-------------------------------------------

The unstable point is marked as O  in the energy surface of the LMG model in Fig.~\ref{Fig:FotocExploration}~(a) and of the Dicke model in Fig.~\ref{Fig:FotocExploration}~(c).  Points O, A, B, and C correspond to the center of the coherent states used in the calculation of the FOTOC. The choices of A, B, and C are done such that the trajectories do not go (come) asymptotically to (from) the unstable point. To guarantee this, since the LMG model has only one degree of freedom, the points A, B, and C have decreasing energies, while for the Dicke model, it is enough to select different values of $Q$ with the same energy $H_D=-\omega_0$.

For any of the points (and for those in between them), the initial evolution of the FOTOC is exactly the same as the one for O, with the same exponential growth rate $2\Lambda \approx 2 \lambda$, as clearly seen in Fig.~\ref{Fig:FotocExploration}(b) [Fig.~\ref{Fig:FotocExploration}(d)] for the LMG [Dicke] model. What changes is the duration of the exponential behavior, which becomes shorter as one gets further from O, and also the saturation value of the dynamics, which gets lower and shows larger oscillations. 

Figure~\ref{Fig:FotocExploration} demonstrates that, in absolute contrast with the classical dynamics, quantum instability is not only possible, but is the rule for generic states in the vicinity of an unstable point. One needs to move quite far from the unstable point to get rid of any reminiscence of an exponential growth. 

{\it Discussion.} Classical systems in the regular regime, as the LMG and the Dicke model considered here, can exhibit unstable points with equal positive classical and quantum LEs. This parallel ceases to hold in the vicinity of the unstable points. Classically, this surrounding area has zero LEs. In the quantum domain, on the other hand, generic states in this region still give positive quantum LEs. Therefore, while one can say that in the vicinity of the unstable points, the quantum-classical correspondence still holds, given that the exact quantum evolution and the TWA match, the same does not hold for the correspondence between the quantum and classical LEs.

Our results are of particular relevance for ongoing experiments with ion traps that aim to investigate quantum chaos in the Dicke model. We show that for quantities, initial states and parameters probed by these experiments, they may eventually detect the effects of unstable points, not necessarily of chaos.

We stress, however, that there is not yet agreement on what quantum chaos really is. If we were to adopt here the simplified and widespread view that it means the exponential growth of OTOCs, we would no longer be able to associate it with the presence of positive classical LEs. Resorting to the more traditional definition of quantum chaos based on level statistics as in random matrix theory does not circumvent the problem either, since Wigner-Dyson distributions have been found also in systems that are classically regular~\cite{Benet2003,Relano2004}. The question ``What are the unquestionable signatures of classical chaos in the quantum domain?'' remains open.

\begin{acknowledgments}
We thank J. Dale for helping us with the proof of Eq.~(S4) in the SM~\cite{SM}, and E. Palacios, L. D\'\i az, and E. Murrieta of the Computation Center-ICN for their support. M.A.B.M. is grateful to J. D. Urbina and K. Richter for their hospitality and the opportunity for exchange of ideas. P.S. is grateful to P. Cejnar for stimulating discussions. We acknowledge financial support from Mexican CONACYT project CB2015-01/255702 and DGAPA, UNAM project IN109417. P.S. is supported by the Charles University Research Center UNCE/SCI/013. L.F.S. is supported by NSF Grant No. DMR-1603418. L.F.S. and J.G.H. acknowledge the hospitality of the Aspen Center for Physics and the Simons Center for Geometry and Physics at Stony Brook University, where some of the research for this Rapid Communication was performed.
\end{acknowledgments}

%%%%%%%%%%%%%%%%% BIBLIO %%%%%%%%%%%%%%%%%%%%%%%
%\bibliography{BIB_biblioinstability}

\begin{thebibliography}{102}
\bibitem [{foo({\natexlab{b}})}]{footRegular}%
  \BibitemOpen
  \bibinfo {note} {Exceptions are the
  orbits asymptotically going to or coming from the point.}\BibitemShut {Stop}%
\end{thebibliography}

%%%%%%%%%%%%%%%%%%%%% END %%%%%%%%%%%%%%%%
 \end{document}

\title{Supplemental Material: Positive quantum Lyapunov exponents in experimental systems with a regular classical limit}
\author{Sa\'ul Pilatowsky-Cameo}
\affiliation{Instituto de Ciencias Nucleares, Universidad Nacional Aut\'onoma de M\'exico, Apdo. Postal 70-543, C.P. 04510  CDMX, M\'exico}
\author{Jorge Ch\'avez-Carlos}
\affiliation{Instituto de Ciencias Nucleares, Universidad Nacional Aut\'onoma de M\'exico, Apdo. Postal 70-543, C.P. 04510  CDMX, M\'exico}
\author{Miguel A. Bastarrachea-Magnani}
\affiliation{Department of Physics and Astronomy, Aarhus University, Ny Munkegade, DK-8000 Aarhus C, Denmark.}
\author{Pavel Str\'ansk\'y}
\affiliation{Faculty of Mathematics and Physics, Charles University, V Hole\v{s}ovi\v{c}k\'ach 2, Prague 180 00, Czech Republic}
\author{Sergio Lerma-Hern\'andez}
\affiliation{Facultad  de F\'\i sica, Universidad Veracruzana, Circuito Aguirre Beltr\'an s/n, C.P. 91000  Xalapa, Mexico}
\author{Lea F. Santos}
\affiliation{Department of Physics, Yeshiva University, New York, New York 10016, USA.} 
\author{Jorge G. Hirsch} 
\affiliation{Instituto de Ciencias Nucleares, Universidad Nacional Aut\'onoma de M\'exico, Apdo. Postal 70-543, C.P. 04510  CDMX, M\'exico}

%%%%%%%%%%%%%%%%%%%%%%%%%%%%%%%%%%%%%%%%

\maketitle

We provide here mathematical details and additional figures for the results discussed in the main text.

%%%%%%%%%%%%%%%%%%%%%%%%%%%%%%%%%%%%%%%%%%%%%%%%
\section{Lyapunov exponent for stationary points of classical Hamiltonian systems } 
\label{app:lyapunov_exponents}

In this section we show how to obtain the Lyapunov exponents (LEs) associated with the stationary points of classical systems.
Consider a classical Hamiltonian system $H(\bm{x})$ with $n$ degrees of freedom and a set of generalized coordinates and respective momenta $\bm{x}=(\bm{q},\bm{p})=(q_1,...,q_n,p_1,...p_n)$. We denote the Hamilton dynamical equations by $\dot{\bm{x}}=\bmcal{F}(\bm{x})$. A stationary point $\bm{x}_0$ of the system satisfies $\bmcal{F}(\bm{x}_0)=0$, {\em i.e.} $\bm{x}_0(t) = \bm{x}_0(t_0) $.  

To calculate the LE $\lambda$ associated with $\bm{x}_0$, we employ the tangent space by means of the fundamental matrix of the system, $\boldsymbol{\Phi}_{\bm{x}}(t)$~\cite{Gaspard1998}. This matrix solves the simultaneous equations
\begin{align*}
\begin{pmatrix}
\dot{\bm{x}} \\
\dot{\boldsymbol{\Phi}}_{\bm{x}} \\
\end{pmatrix}=\begin{pmatrix}
\bmcal{F}(\bm{x}) \\
D_{\bm{x}}\bmcal{F}(\bm{x})\boldsymbol{\Phi_{\bm{x}}}\\\end{pmatrix}, &&&&
\begin{pmatrix}
\bm{x}(t_{0}) \\
\boldsymbol{\Phi_{\bm{x}}}(t_{0}) \\
\end{pmatrix}=\begin{pmatrix}
\bm{x}_{0} \\
\mathbb{I}_{2n}\\\end{pmatrix},
\end{align*}
where $D_{\bm{x}}\bmcal{F}(\bm{x})$ is the Jacobian matrix of $\bmcal{F}$ and $\mathbb{I}_{2n}$ is the ${2n \times 2n}$ identity matrix. Because $\bm{x}_0$ is a stationary point, 
$A=D_{\bm{x}}\bmcal{F}(\bm{x})=D_{\bm{x}}\bmcal{F}(\bm{x}_0)$ is independent of time and
%. 
\begin{equation}
\boldsymbol{\Phi}_{\bm{x}_{0}}(t)=e^{At}.
\end{equation}
The fundamental matrix allows us to find the time evolution of a variation over the initial condition $ \delta\bm{x}_{0}$. Specifically,
\begin{equation}
\delta\bm{x}(t)=\boldsymbol{\Phi}_{\bm{x}_{0}}(t)\,\delta\bm{x}_{0}=e^{At}\,\delta\bm{x}_{0}.
\end{equation}
Employing the spectral norm $\Vert \boldsymbol{\Phi}_{\bm{x}_{0}}(t)\Vert$~\cite{horn_johnson_1985} 
(the square root of the largest eigenvalue of the matrix $\boldsymbol{\Phi}_{\bm{x}_{0}}^{\dagger} \boldsymbol{\Phi}_{\bm{x}_{0}} $), one gets \cite{Gaspard1998},
\begin{equation}
\lambda=\lim_{t\rightarrow\infty}\;\dfrac{1}{t} \log \Vert e^{At}\Vert.
\label{ec:LCET}
\end{equation}
This limit can be expressed in terms of the eigenvalues $\lambda_i$ of the matrix $A$. Denoting the maximum of the real parts of the eigenvalues of $A$ by ${\lambda_{\max}(A)=\max_i{\Re \left(\lambda_i \right)}}$, we have that
\begin{equation}
\lambda=\lambda_{\text{max}}(A) .
\label{eq:LCE_Eigenformula}
\end{equation}

%%%%%%%%%%%%%%%%%%%%%%%%%%%%%%%%%%%%%%%%
\begin{proof}
\newcommand{\mmatrix}[1]{\mathrm{#1}}

This proof was modeled after a discussion in~\cite{DaleStackExchange2019}. 

First, note that both sides of Eq.~\eqref{eq:LCE_Eigenformula} are invariant under arbitrary basis changes. 
For any ${A'=PAP^{-1}}$, with $P$ being any invertible complex matrix, as both $A$ and $A'$ share the same eigenvalues, on the right hand side of Eq.~\eqref{eq:LCE_Eigenformula}, $\lambda_{\text{max}}(A)=\lambda_{\text{max}}(A')$. On the other side of Eq.~\eqref{eq:LCE_Eigenformula}, using the submultiplicativity of the spectral norm, we have 
\begin{equation*}
\left|\log \Vert e^{At}\Vert-\log\Vert e^{A't}\Vert\right| \leq \log(\Vert P\Vert \Vert P^{-1}\Vert),
\end{equation*}
and then
\begin{equation*}
\lim_{t \to \infty} \frac{1}{t }\log \Vert e^{At} \Vert=\lim_{t \to \infty} \frac{1}{t }\log \Vert e^{A't} \Vert.
\end{equation*}

Because of the above, we can write $A$
in Jordan normal form ${A=\text{diag}(J_1,J_2, ... , J_N)}$, where each Jordan block $J_i$ corresponds to an eigenvalue $\lambda_i$ of $A$ with multiplicity $n_i$ and is given by ${J_i=\lambda_i \mathbb{I}_{n_i}+ S_{n_i}}$. Here, $S_{n_i}$ is the ${n_i\times n_i}$ matrix with superdiagonal elements equal to one and zero elsewhere. We then get
\begin{equation*}
\norm{e^{At}}=\norm{\text{diag}(e^{J_1 t},e^{J_2 t}, ... , e^{J_N t})}=\max_i\norm{ e^{J_i t}},
\end{equation*}
so 
\begin{equation}
\lim_{t \to \infty} \frac{1}{t }\log \norm{e^{At}}=\max_i\left(\lim_{t \to \infty} \frac{1}{t }\log\norm{e^{J_i t}} \right).    
\label{eqn:limitasmaximumofjordanblocks}
\end{equation}

Since the matrices $\lambda_i \mathbb{I}_{n_i}$ and $S_{n_i}$ commute,
$$e^{J_i t}=e^{\lambda_i \mathbb{I}_{n_i} t}e^{S_{n_i} t}=e^{\lambda_i t} \sum_{k=0}^{n_i-1} \frac{t^k}{k !} S_{n_i}^k,$$
with spectral norm 
$$\norm{ e^{J_i t}}=e^{\Re(\lambda_i) t} \norm{\sum_{k=0}^{n_i-1} \frac{t^k}{k !} S_{n_i}^k}.$$
Thus,
\begin{align}
    \label{eqn:limitofjordanblockextra}
    \lim_{t \to \infty} \frac{1}{t }\log\norm{ e^{J_i t}} =&\Re(\lambda_i)  \\& + \lim_{t \to \infty} \frac{1}{t }\log\norm{\sum_{k=0}^{n_i-1} \frac{t^k}{k !} S_{n_i}^k}. \nonumber
\end{align}

To complete the proof we
have to show that the last term in Eq.~\eqref{eqn:limitofjordanblockextra} equals zero. If ${n_i=1}$, it is trivial, so assume ${n_i>1}$. Using the triangle inequality of the norm and the fact that $\norm{ S_{n_i}^k}=1$ for all ${k\in \left \{0,1,...,n_i-1 \right\}}$, we find the bounds
\begin{equation} p_L(t) \leq \norm{\sum_{k=0}^{n_i-1} \frac{t^k}{k !} S_{n_i}^k} \leq p_R(t),
    \label{eqn:leftandrightpbounds}
\end{equation}
where
\begin{align*}
    p_L(t)=\frac{t^{n_i-1}}{(n_i-1)!}-\sum_{k=0}^{n_i-2} \frac{t^k}{k !}, && p_R(t)=\sum_{k=0}^{n_i-1} \frac{t^k}{k !}.
\end{align*}
For large enough $t$, $p_L(t) >0$, and we may take the logarithm and divide by $t$ the three terms in Eq. \eqref{eqn:leftandrightpbounds}. But for any real polynomial $p(t)$ with positive leading coefficient, one has $\lim_{t\to \infty}\frac{\log p(t)}{t}=0$ by applying  L'H\^opital's rule. Thus, using the squeeze theorem in Eq.~\eqref{eqn:leftandrightpbounds}, we get
\begin{equation}
\lim_{t \to \infty} \frac{1}{t }\log \norm{\sum_{k=0}^{n_i-1} \frac{t^k}{k !} S_{n_i}^k}=0.
\label{eqn:limitofextraequalszero}
\end{equation}
With Eq.~$\eqref{eqn:limitasmaximumofjordanblocks}$, Eq.~$\eqref{eqn:limitofjordanblockextra}$, and Eq.~$\eqref{eqn:limitofextraequalszero}$, we arrive at the desired result,
\begin{equation}
\lim_{t \to \infty} \frac{1}{t }\log \norm{e^{At}}=\max_i\Re(\lambda_i) = \lambda_{\text{max}}(A).    
\end{equation}
Notice that we did not use any particular property of the matrix $A$ at any point. This formula is true for any complex matrix $A$.
\end{proof}

In what follows we apply this result to obtain the LEs for the classical simple pendulum, the Lipkin-Meshkov-Glick (LMG) model, and the Dicke model.

%%%%%%%%%%%%%%%%%%%%%%%%
\subsection{Lyapunov exponent of the simple pendulum} 
The Hamiltonian of the classical simple pendulum is
\begin{equation}
H(\theta,p_\theta)=\dfrac{p^2_\theta}{2ml^2}+mgl(1-\cos\theta),
\end{equation}
where $m$ and $l$ are the mass and length of the pendulum, respectively, $g$ is the gravity acceleration, $\theta$ is the angle of the pendulum measured from the vertical, and $p_\theta=m l^2 \dot{\theta}$ is the canonical momentum associated with $\theta$.
The Hamilton equations are $\dot{\bm{x}}=\bmcal{F}(\bm{x})=(\partial_{p_\theta} H, -\partial_\theta H)$ with $\bm{x}=(\theta,p_\theta)$. This model has two stationary points, $\bm{x}_{\downarrow}=(0,0)$ at the bottom and $\bm{x}_{\uparrow}=(\pi,0)$ at the top. The Jacobian matrix for each point can be obtained by simple differentiation, 
\begin{align*}
A= \frac{1}{m l^2}\begin{pmatrix} 0 & 1 \\ \mp \omega^2 & 0 
\end{pmatrix} ,
\end{align*}
\noindent
where $\omega=\sqrt{g/l}$, the negative sign corresponds to $\bm{x}_{\downarrow}$ and the positive sign to $\bm{x}_{\uparrow}$. 
For $\bm{x}_{\downarrow}$, the eigenvalues of $A$ are pure imaginary numbers. From Eq.~\eqref{eq:LCE_Eigenformula}, $\bm{x}_{\downarrow}$ is then a center point with zero LE, $\lambda_{\downarrow}=0$, that is, it is a stable point. For $\bm{x}_{\uparrow}$, the eigenvalues of $A$ are $\pm \omega$. This is a hyperbolic point with $\lambda_{\uparrow}=\omega$, so it is an unstable point. Despite being the go-to example of an integrable system, even the simple pendulum has a non-zero LE at $\bm{x}_{\uparrow}$. 

In addition, this positive LE is also expressed in the orbits emanating from the instability. These orbits, comprised of all initial conditions with the same energy as $\bm{x}_{\uparrow}$, asymptotically go to the unstable point. When one computes $\lambda$ on such points, one obtains the same positive LE. 

All other initial conditions, outside the orbits that emanate from the instability, have a zero LE. They give rise to closed periodic orbits, and because the pendulum is an integrable system, such orbits must be stable, with a null LE \cite{HilbornBook,PikovskijBook}. Therefore, the region of positive LE has measure zero and the system cannot be classified as chaotic.

 %%%%%%%%%%%%%%%%%%%%%%%%
 \subsection{Lyapunov exponent of the Lipkin-Meshkov-Glick model} 

To find the stationary points of the LMG model, we use the equations of motion for $H_{\text{LMG}}$ (see this Hamiltonian in Eq.~(2) of the main text), which are 
\begin{align}
   &\label{eqn:dotQ}\Dot{Q}=\quad \frac{\partial H_{\text{LMG}}}{\partial P }=  P \left(\Omega-\frac{\xi Q^2}{2}\right) \\
   &\label{eqn:dotP}\Dot{P}=  -\frac{\partial H_{\text{LMG}}}{\partial Q }=Q \left(\frac{\xi P^2}{2}-(2 \xi+\Omega)\right)+\xi Q^3.
\end{align}
The stationary point happens at $\bm{x}_0=(Q,P)=(0,0)$,
where the Jacobian matrix is
\begin{equation}
 \label{Eq:LipA}
  A=\begin{pmatrix}
    0 & \Omega \\
    -(\Omega+ 2\xi) & 0
    \end{pmatrix}.
\end{equation}
Its eigenvalues are $\lambda_{\pm}=\pm\sqrt{-(\Omega^2+2\xi\Omega)}$. According to Eq.~\eqref{eq:LCE_Eigenformula}, the LE for this stationary point is zero for $\Omega\geq -2\xi$, because all the eigenvalues of $A$ are imaginary. However, for $\Omega< -2\xi$, both eigenvalues of $A$ are real and the LE equals the positive one. We then have~\cite{Pappalardi2018}
\begin{align}\label{Eq:LyLip}
	\lambda = 
		\begin{cases}
		0 & \text{if } \Omega\geq -2\xi \\ 
		\sqrt{-(\Omega^2+2 \Omega\xi)} & \text{if } \Omega< -2\xi
		\end{cases}.
\end{align}

Just as in the simple pendulum, because the LMG model has one degree of freedom and conserves energy, it is integrable. Thus, all points other than the unstable stationary point and the trajectories that emanate from it give rise to stable periodic orbits with zero LE~\cite{HilbornBook,PikovskijBook}.

 %%%%%%%%%%%%%%%%%%%%%%%%
 \subsection{Lyapunov exponent of the Dicke model}

The Dicke model has a ground-state quantum phase transition (QPT)~\cite{Kloc2017,Bastarrachea2016JSTAT} at 
\begin{gather}
\omega_{0c}=\frac{4 \gamma^2}{\omega}.
\end{gather}
It is in the normal phase when $\omega_0 > \omega_{0c}$ and in the superradiant when $\omega_0 < \omega_{0c}$.

The Hamilton equations of the Dicke model have several stationary points depending on the set of Hamiltonian parameters that are chosen (see details in Refs.~\cite{Bastarrachea2014a,Bastarrachea2016JSTAT}). We are interested in the unstable stationary point, which is related with the excited state quantum phase transition (ESQPT). 
In the classical phase space, this point is $\bm{x}_0=(Q,q,P,p,)=(0,0,0,0)$, which, according to the classical Hamiltonian given in Eq.~(4) of the main text, implies energy $-\omega_0$. In the normal phase, this point is stable and corresponds to the quantum ground state in the thermodynamic limit. Keeping $\omega$ and $\gamma$ of the Dicke Hamiltonian constant (see the main text), this point becomes unstable if one decreases the atomic level spacing, so that $\omega_{0}<\omega_{0c}$ \cite{Kloc2017}.

The Jacobian matrix at $\bm{x}_0$ is
\begin{equation}
A=\begin{pmatrix}
0 & 0 & \omega_0 & 0   \\
0 & 0 & 0 & \omega \\
-\omega_0 & -2\gamma & 0 & 0 \\
-2\gamma & -\omega & 0 & 0
\end{pmatrix},
\end{equation}
and has the following eigenvalues
\begin{equation}
\lambda_{1,2,3,4} =\pm\dfrac{1}{\sqrt{2}}\sqrt{-\left(\omega^2+\omega^2_0\right)\pm\sqrt{\left(\omega^2-\omega^2_0\right)^2+ 16\gamma^2 \omega \omega_0}}.
\end{equation}
If $\omega_{0}>\omega_{0c}$, all eigenvalues are imaginary and the LE is zero, but if $\omega_{0}<\omega_{0c}$, some eigenvalues become real, and the largest one is given by
\begin{equation}
\lambda=\frac{1}{\sqrt{2}} \sqrt{-\left(\omega^2 + \omega_0^2\right) + \sqrt{\left(\omega^2-\omega^2_0\right)^2+ 16\gamma^2 \omega \omega_0}}.
\label{Eq:LyD}
\end{equation}

Points different from the unstable point and the orbits that emanate from it may have positive LEs, as the Dicke model is chaotic for some energy regimes. For a detailed analysis, see \cite{Chavez2019}. In this work, however, our focus is on the regular regime.

%%%%%%%%%%%%%%%%%%%%%%%%%%%%%%%%%%%%%%%%
\section{Fidelity out-of-time order correlator} 
\label{app:FOTOC}

The out-of-time order correlator (OTOC) is defined as
\begin{equation}
F(t)= \langle \hat{W}^\dagger (t) \hat{V}^\dagger  \hat{W} (t) \hat{V}\rangle ,
\label{Eq:fotoc}
\end{equation}
where
$\hat{W} (t) = e^{i \hat{H} t} \hat{W} e^{-i \hat{H} t}$. If $ \hat{W}$ and $\hat{V}$ are both unitary operators, 
\begin{equation}
 \Re(F(t)) = 1 - \langle \left[\hat{W},\hat{V}\right]^\dagger \left[\hat{W},\hat{V}\right] \rangle /2.
 \label{eq:ReF}
\end{equation}

In Ref.~\cite{Lewis-Swan2019}, the fidelity OTOC (FOTOC) is obtained by using as operators 
 \begin{equation}
 \hat{W}= e^{i \delta\phi \hat{G}}, 
 \label{eq:W}
 \end{equation}
where $\hat{G}$ is a Hermitian operator and $\delta\phi$ is a perturbative small parameter, and
 \begin{equation}
 \hat{V} = | \Psi_0\rangle \langle   \Psi_0| ,
 \end{equation}
which is the projector onto the initial state $| \Psi_0\rangle$.
Using the following expansion up to second order in $\delta\phi$,
\begin{equation}
\langle \Psi_0| \hat{W}(t) | \Psi_0\rangle  \approx 1 + i  \delta\phi \langle \hat{G}(t) \rangle- \frac{\delta\phi^2}{2} \langle \hat{G}(t)^2 \rangle,
\end{equation}
 it can be shown that $F(t)$ is related with the variance of $\hat{G}$ as
 \begin{equation}
\frac{1 - F(t)}{(\delta\phi)^2}  \approx  ( \langle \hat{G}(t)^2 \rangle -\langle \hat{G}(t) \rangle^2 ) \equiv \sigma^2_{G}(t) .
\end{equation}
Since the evolution of $F(t)$ is equivalent to that of $\sigma^2_{G}(t) $, we refer to the variance as FOTOC.

For most interacting many-body systems, the OTOCs must be calculated numerically. To do so for time-independent Hamiltonians, one employs the eigenbasis expansion. The FOTOC is then calculated as 
\begin{eqnarray}
\sigma^2_{\hat{G}}(t)=\sum_{i,j,k} c_j^{0 *} c_i^0 e^{i(E_j -E_i) t} G_{jk} G_{ki} \nonumber \\ - \left(\sum_{i,j} c_j^{0 *} c_i^0 e^{i(E_j -E_i) t} G_{ji}\right)^{2},
\label{eq:GG}
\end{eqnarray}
where $G_{ij}=\langle E_i |\hat{G}|E_j\rangle$,  the energy eigenstate is $|E_i\rangle$, and $c^0_i=\langle \Psi_{0}|E_i\rangle$. Even though Eq.~(\ref{eq:GG}) is formally simple, it is, in general, numerically challenging for non-integrable systems, such as the Dicke model. In this case, we use exact quantum evolution when the system size is not too large and to deal with large sizes, we resort to the truncated Wigner approximation (TWA), which is a phase space method.

%%%%%%%%%%%%%%%%%%%%%%%%%%%%%%%%%%%%%%%%
\section{Truncated Wigner approximation for the Dicke Model}
\label{app:coherent_wigner}

The calculation of the FOTOC, $\sigma_{x_i}^2=\langle \hat{x}_i^2 \rangle-\langle \hat{x}_i \rangle^2$, is reduced to obtaining the expectation value of powers of the canonical coordinates $\hat{x}_{i}$ of a quantum system under the short time evolution of a certain quantum state $|\Psi_0\rangle$. To do this, we use the TWA.

Consider the Wigner function~\cite{Wigner1932} of the evolved initial state, $W(\bm{x},t)$. The expectation value of an observable $\hat{O}$ may be calculated by
\begin{equation*}
\langle \hat{O} (t) \rangle=\int W(\bm x,t) O(\bm x)  \text{d} \bm x ,
\end{equation*} 
where $O(\bm x)$ is a function of the phase space variables known as the Weyl symbol of $\hat{O}$ \cite{Weyl1927}.
In our case, the observables of interest are  $\hat{O}=\hat{x}_{i}^n$ with $n=1,2$. Their Weyl symbols, in very good approximation (the error is of order $j^{-1}$), are obtained by removing the hats off the observable $O=x_{i}^n$.
Thus,
\begin{equation}
    \label{eqn:Wig_expvalues}
    \left \langle \hat{x}_i^n(t) \right \rangle  = \int W(\bm{x},t) x_i^n \text{d} \bm x.
\end{equation}

The TWA is used to deal with the evolution of the Wigner function~\cite{Steel1998,Blakie2008}. The idea of the approximation is to treat a positive Wigner function as a classical phase space distribution for short times, assuming that it remains constant over the classical phase space trajectories. 
The temporal evolution is then given by~\cite{Gorin2006}
\begin{equation} 
    W(\bm{x},t) = W(\bm x(-t),0),
\end{equation}
where $\bm x(t)$ is the classical trajectory in the classical Hamiltonian phase space. Inserting this in Eq.~\eqref{eqn:Wig_expvalues}, we get
\begin{equation} 
\label{eq:semiclassicalexpectedvalue}
    \left \langle \hat{x}_i^n (t) \right \rangle = \int W(\bm{x}) \,  x_i(t)^n \,\text{d} \bm x.
\end{equation}

%%%%%%%%%%%%%%%%%%%%%%%%%%%%%%%%%%%%
\subsection{Dicke model: TWA-quantum correspondence}

For the Dicke model, we use the product of Glauber $|\alpha\rangle$ and Bloch $|z\rangle$ coherent states of the Heisenberg-Weyl and the SU(2) spaces, respectively, as initial states ~\cite{Zhang1990}. The Wigner function for these states is everywhere positive.

For Glauber states, the Wigner function is given by a normal distribution~\cite{Case2008}
\begin{equation} 
\label{eqn:Wqp_def}
W_{q_0,p_0}(q,p)=\frac{j}{\pi}e^{-j \Delta^2},
\end{equation}
where $\Delta=\sqrt{\left(q-q_0 \right)^2 + \left(p-p_0 \right)^2} $ is the distance between $(q,p)$ and $(q_0,p_0)$.

For a Bloch coherent state, the Wigner function may be written as a sum of Legendre polynomials $P_k(x)$ \cite{typoRef,Klimov2019}
\begin{equation} \label{eq:wigbloch} 
W_{\theta_0,\phi_0}(\theta,\phi)=\frac{(2j)!}{4\pi}\sum_{k=0}^{2j}\sqrt{\frac{(2k+1)}{(2j-k)!(2j+k+1)!}} P_k (\cos \Theta), 
\end{equation}
where $z=\tan(\theta/2)e^{i\phi}$ and $\Theta$ is the angle between $(\theta,\phi)$ and $(\theta_0,\phi_0)$ obtained from
$$\cos\Theta = \cos \theta \cos \theta_0 + \sin \theta \sin \theta_0 \cos(\phi - \phi_0).$$
As $j$ increases, Eq.~\eqref{eq:wigbloch}  converges rapidly to a normal distribution on the Bloch sphere
\begin{equation} 
\label{eqn:WQP_def}W_{\theta_0,\phi_0}(\theta,\phi)\approx\frac{j}{\pi}e^{-j \Theta^2}.
\end{equation}

\vskip 0.5 cm
\begin{figure}[t]
\centering
\begin{tabular}{c}
\includegraphics[width= 0.45 \textwidth]{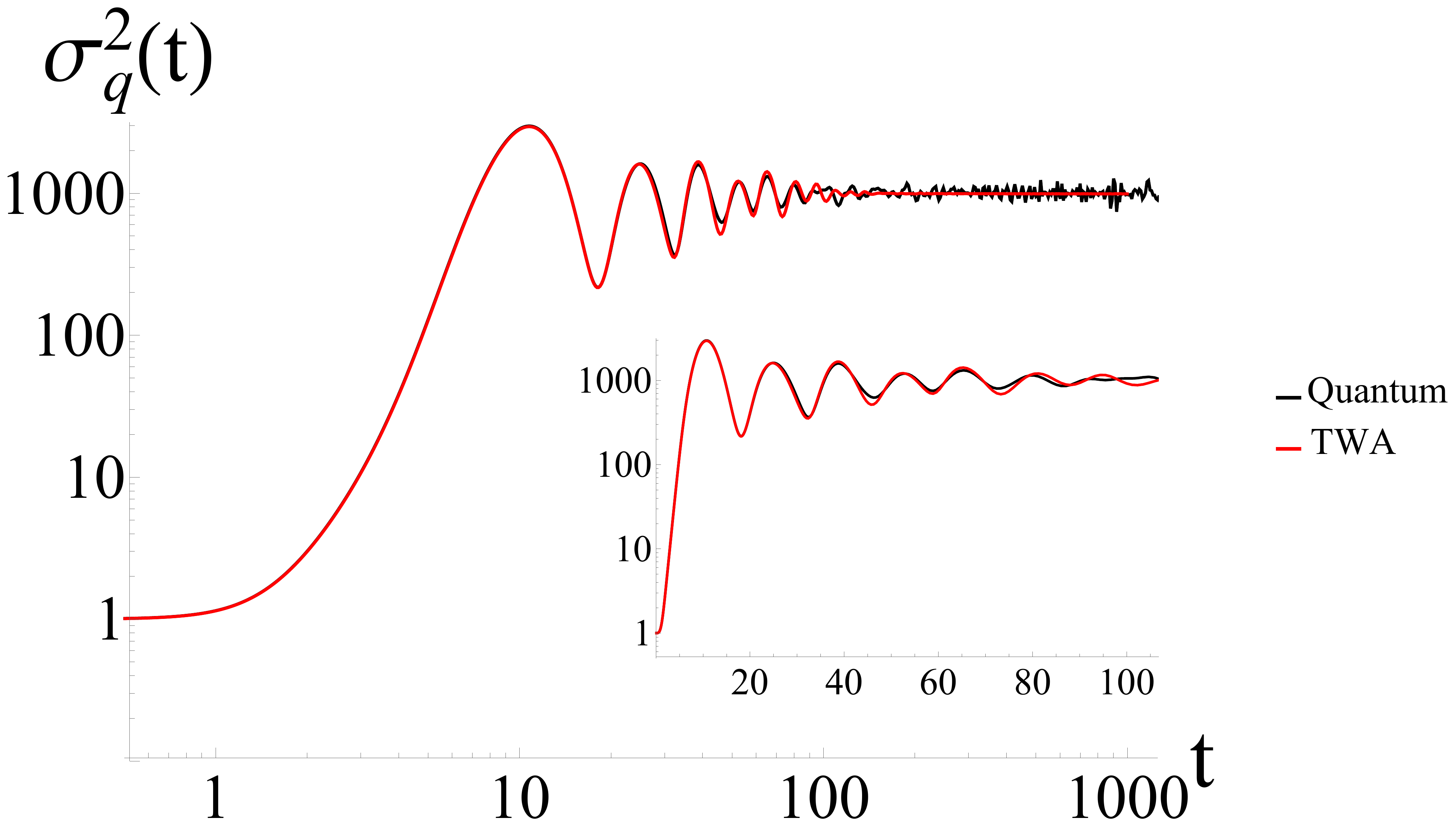} \\
(a) \\
\includegraphics[width= 0.45 \textwidth]{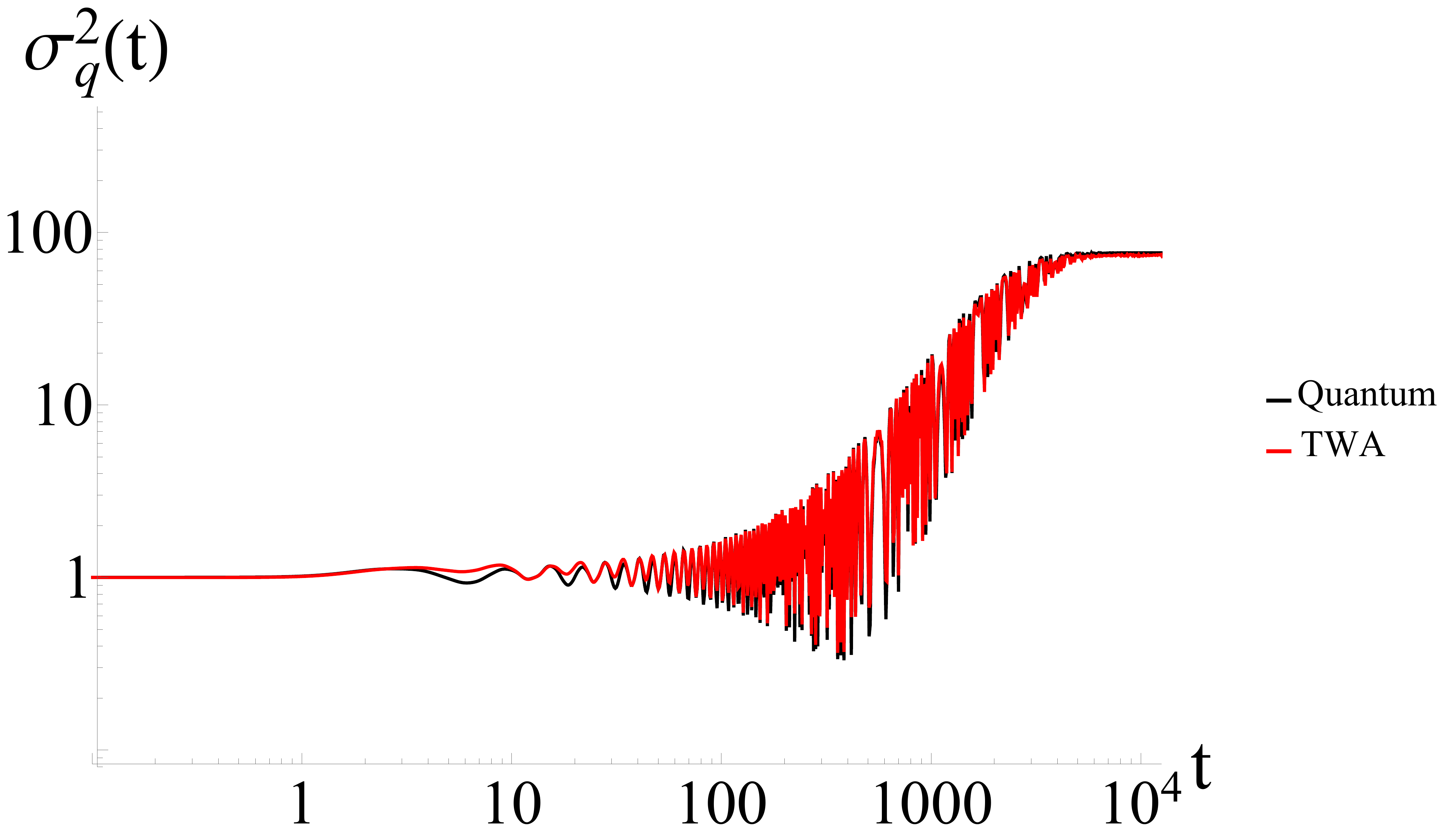} \\
(b)
\end{tabular} 
\caption{\small (Color online) FOTOC for the observable $\hat{q}$ in the Dicke model. The coherent state has (a) energy $E= -\omega_0 j$ centered at the unstable stationary point $\bm{x}_0=(q=0,p=0,Q=0,P=0)$ and (b) $E=-2.5j\omega_0$ centered in a regular point $\bm{x}=(q=-1.722,p=0,Q=1.2,P=0)$. In both panels, the parameters are the experimental values used in Ref.~\cite{Lewis-Swan2019,Safavi2018}: $\omega_0=0.649$, $\omega=0.5$, $\gamma=0.66$, and $j=100$.  The black line corresponds to the actual quantum evolution, while the red line is the truncated Wigner approximation. The inset in panel (a) shows the same data, but in a lin-log scale.}
 \label{Fig:Fotoc_Dicke_QvsSC}
\end{figure}

The Wigner function of a Dicke coherent state centered at $\bm{x_0}$ is given by the product of the Glauber and Bloch Wigner functions
\begin{equation} 
\label{eqn:W_total_def}
W_{\bm{x_0}}(\bm x)=\left (\frac{j}{\pi}\right )^2e^{-j\left(\Delta^2 +\Theta^2  \right)}.
\end{equation}
To compute the integral in Eq.~\eqref{eq:semiclassicalexpectedvalue} with this initial Wigner function, we use a Monte Carlo method~\cite{Schachenmayer2015}, where we sample a set of $\sim 10^4$ initial conditions $\bm x$ from the initial normal Wigner distribution in Eq.~\eqref{eqn:W_total_def} and then calculate the mean of the values of $x_i ^n$ after evolving the points according to the classical Hamiltonian.    The approximation in Eq.~\eqref{eqn:WQP_def}
allows us to simplify the sampling from the distribution and save computational resources.

In Fig.~\ref{Fig:Fotoc_Dicke_QvsSC}, we compare the temporal TWA evolution with the exact quantum evolution in the Dicke model to determine where they diverge. We calculate the FOTOC of the observable $\hat{q}$ for a coherent state located at the critical point $\bm{x}_0$ for a given value of $\omega_0$. The results are shown in Fig.~\ref{Fig:Fotoc_Dicke_QvsSC} (a). Notice that there is perfect agreement for times beyond the exponential growth of the variance, which provide us with a solid foundation to
use this approximation to analyze the short time quantum behavior of the FOTOC for large system sizes, as we do in the main text. 

Interestingly, the semiclassical approximation holds for even longer times in the regular case. This is seen in Fig. \ref{Fig:Fotoc_Dicke_QvsSC} (b), where we compare the TWA and the exact quantum evolution for a coherent state located in a regular region of the phase space. In this case, we select a lower energy $E/j\omega_{0}=-2.5$ and a point with ${\bm{x}=(q=-1.722,p=0,Q=1.2,P=0)}$. We choose this point because it is located roughly at the center of the available phase space that is completely regular at this energy.  The agreement  between the semi-classical and quantum simulations lasts for a long time. 

%%%%%% REFERENCCES %%%%%%%%%
%\bibliography{BIB_biblioinstability} 
